\documentclass[lettersize,journal]{IEEEtran}
\usepackage{amsmath,amsfonts}
\usepackage{algorithmic}
\usepackage{algorithm}
\usepackage{array}
\usepackage[caption=false,font=normalsize,labelfont=sf,textfont=sf]{subfig}
\usepackage{textcomp}
\usepackage{stfloats}
\usepackage{url}
\usepackage{verbatim}
\usepackage{graphicx}
\usepackage{cite}
\usepackage[table]{xcolor}
\usepackage{booktabs}
\usepackage{multirow}
\usepackage{makecell}
\usepackage{threeparttable}
\usepackage[normalem]{ulem}
\usepackage{flushend}
\usepackage{soul,color,hyperref,listings}
\soulregister\cite7
\soulregister\citep7
\soulregister\citet7
\soulregister\ref7
\soulregister\pageref7
\soulregister\em7

\begin{document}

\title{Toward Intelligent and Secure Cloud: Large Language Model Empowered Proactive Defense}

\author{Yuyang~Zhou,~\IEEEmembership{Member,~IEEE},
~Guang~Cheng,~\IEEEmembership{Member,~IEEE},
~Kang~Du,
~Zihan~Chen,
~and Yuyu~Zhao
\thanks{The authors are with the School of Cyber Science and Engineering, Southeast University, Purple Mountain Laboratories, and Jiangsu Province Engineering Research Center of Security for Ubiquitous Network, Nanjing 211189, China.}
\thanks{Guang Cheng is the corresponding author.}}

\markboth{Journal of \LaTeX\ Class Files,~Vol.~14, No.~8, August~2021}%
{Shell \MakeLowercase{\textit{et al.}}: A Sample Article Using IEEEtran.cls for IEEE Journals}


\maketitle

\begin{abstract}
  The rapid evolution of cloud computing technologies and the increasing number of cloud applications have provided numerous benefits in our daily lives. However, the diversity and complexity of different components pose a significant challenge to cloud security, especially when dealing with sophisticated and advanced cyberattacks such as Denial of Service (DoS). Recent advancements in the large language models (LLMs) offer promising solutions for security intelligence. By exploiting the powerful capabilities in language understanding, data analysis, task inference, action planning, and code generation, we present LLM-PD, a novel defense architecture that proactively mitigates various DoS threats in cloud networks. LLM-PD can efficiently make decisions through comprehensive data analysis and sequential reasoning, as well as dynamically create and deploy actionable defense mechanisms. Furthermore, it can flexibly self-evolve based on experience learned from previous interactions and adapt to new attack scenarios without additional training. Our case study on three distinct DoS attacks demonstrates its remarkable ability in terms of defense effectiveness and efficiency when compared with other existing methods.
\end{abstract}


\section{Introduction}
\IEEEPARstart{C}{loud} computing has accelerated rapidly in recent years, evolving into a cornerstone technology for modern communication systems such as 5G/6G networks, the Internet of Things (IoT), and edge computing. The integration of cloud-native technologies with Network Functions Virtualization (NFV) and Software-Defined Networking (SDN) is transforming how communication services are deployed, managed, and secured. As a result, the security and resilience of cloud platforms directly impact the reliability and performance of contemporary communication infrastructures.

Despite these advantages, the diversity and complexity of cloud components, i.e., networks, architectures, Application Programming Interfaces (APIs), and hardware, has posed significant security challenges, especially for communication systems that demand high reliability. The use of standard Internet protocols and virtualization technologies increases the attack surface, making cloud-based communication infrastructures susceptible to a range of threats, e.g., IP spoofing and Denial of Service (DoS) attacks. Moreover, emerging threats like zero-day vulnerabilities and Advanced Persistent Threats (APTs) present additional challenges that traditional solutions may not effectively address, particularly in dynamic and large-scale environments.

To address the aforementioned challenges, several proactive defense techniques have been proposed, including Moving Target Defense (MTD)~\cite{tan2023survey}, cyber deception~\cite{javadpour2024comprehensive}, Mimic Defense~\cite{Wu2020}, among others. These methods emphasize the proactive identification, warning, and response to potential threats through automated and adaptive mechanisms, either before or during an attack, thereby effectively reducing security risks and minimizing potential losses.

Although these solutions overcome, to some extent, the shortcomings of traditional solutions, they require modifications to mitigation mechanisms that may not be effective across diverse environments. Moreover, the decision-making of defense deployment predominantly relies on heuristic, Machine Learning (ML), and Deep Learning (DL) algorithms. Nevertheless, given the increasing complexity of cloud-based applications and the wide array of attack vectors, it is hard for any specific strategy to fully adapt to the different and time-varying scenarios in the cloud. For this reason, there is an urgent requirement for an intelligent and adaptable approach to facilitate proactive defense within the cloud environment.

Fortunately, Large Language Models (LLMs) have profoundly impacted research and practice~\cite{maatouk2025large}, excelling at in-context learning, prompt-driven reasoning, decision-making, and scenario simulation~\cite{huang2025large}. Building upon these advantages, recent research has leveraged LLMs in the cybersecurity community, empowering security professionals to explore various attack vectors (e.g., vulnerability detection) and develop autonomous agents (e.g., code fixing). Therefore, the extensive knowledge encoded in LLMs has sparked our interest in exploring their potential for protection in the complex and ever-changing cloud security scenarios.

In this paper, differing significantly from the previous defense using expert knowledge or specific strategies, we leverage pre-trained LLMs with different prompts and deploy them across a variety of attack scenarios in the cloud domain, achieving enhanced protection and proactive response. The main contribution of this work includes the following.
\begin{itemize}
  \item We design a novel robust and efficient cloud security architecture called LLM-PD for proactive DoS defense. To the best of our knowledge, this is the first comprehensive end-to-end LLM-driven proactive defense architecture where specialized agents handle different aspects of the defense lifecycle while maintaining contextual awareness in cloud environments.
  \item We achieve strong adaptability across different DoS attack vectors without extensive training, overcoming limitations of traditional ML or DL methods that require retraining for different scenarios. Moreover, we develop a self-evolving memory feedback mechanism that enables continuous learning and improvement of defense capabilities through experience.
  \item We conduct a case study with three LLMs to demonstrate the practical implementation of our architecture and its adaptation to various DoS attacks. Comparative experiments reveal significant advantages in execution accuracy, surviving rate, time efficiency, and efficacy over existing approaches. The source code is available at \url{https://github.com/SEU-ProactiveSecurity-Group/LLM-PD}.
\end{itemize}

The rest of this paper is organized as follows. We commence with the background of proactive defense and discuss the related work of LLM for cybersecurity. Then we elaborate on the proposed architecture with design details introduced. Next, we discuss the prototype construction and use the mitigation of multiple DoS attacks as a case study. We conduct experimental evaluations to illustrate efficient performance. Finally, we conclude the work and analyze future challenges.

\section{Background and Related Work}
\subsection{Proactive Defense in Cloud Networks}
Proactive defense enables anticipatory control over system security, allowing rapid detection of network or behavioral anomalies and timely prevention of suspicious activities. Unlike reactive approaches, it provides a tactical advantage through real-time monitoring and early intervention.

For example, Tuyishime \emph{et al.}~\cite{tuyishime2023enhancing} address security challenges in public cloud environments by proposing a cloud-native Security Information and Event Management (SIEM) system. Their architecture integrates various cloud resources to provide automated visibility and centralized protection. However, such approaches mainly focus on monitoring and alerting, lacking the ability to autonomously generate and adapt defense strategies in timely manner.

Zhou \emph{et~al.}~\cite{zhou2025resource} combine MTD technologies with Deep Reinforcement Learning (DRL) to proactively defend against Low-rate DoS attacks. While this method introduces learning-based adaptation, it requires extensive retraining for each new attack scenario and is limited by the scope of predefined action spaces, making it less flexible in dynamic environments.

Wu \emph{et~al.}~\cite{wu2022intrinsic} propose an intrinsic cloud security (ICS) defense framework that fuses MTD and mimic defense within NFV-based clouds. This integration enhances protection against co-resident attacks, memory DoS, and other resource exhaustion-based threats, but the framework still depends on predefined defense logic and does not generalize well to other threat types.

In summary, existing proactive defense solutions either depend on static rules or are tailored to specific attack types. They generally lack the flexibility, adaptability, and autonomous reasoning required to address the diverse and evolving threats present in cloud environments. In contrast, our proposed architecture leverages the reasoning capabilities of LLMs, and evolves based on the proposed feedback mechanism to enable cross-scenario, end-to-end proactive defense without the need for extensive retraining.

\subsection{LLM for Cybersecurity Enhancement} 
Recent advances in LLMs have transformed cybersecurity by enabling more adaptive and intelligent defense technologies~\cite{yao2024survey}, enhancing tasks such as threat detection, vulnerability analysis, and automated defense mechanisms.

For instance, Shafee \emph{et~al.}~\cite{shafee2024evaluation} compare the performance of both commercial and open-source models on cybersecurity datasets. Their work examines the adaptability of these models to Cyber Threat Intelligence (CTI) tasks. While their findings highlight the flexibility and practical utility of LLMs, the study focuses on information extraction and classification tasks, rather than autonomous defense in the full lifecycle.

Similarly, Levi \emph{et~al.}~\cite{levi2025cyberpal} introduce CyberPal.AI for cybersecurity applications. By leveraging expert-designed schemas and a hybrid data generation process for training, it demonstrates substantial improvements in handling security-related instructions. Although such method significantly improves LLMs' ability to understand and reason about security concepts, its application is limited to knowledge-based tasks.

Additionally, Loevenich \emph{et~al.}~\cite{loevenich2025design} present an Autonomous Cyber Defence (ACD) agent, which combines DRL, augmented LLMs, and rule-based systems to enable automated defensive actions. While this approach demonstrates the effectiveness of LLMs for automated cyber defense, it still relies on a large amount of expert interaction for certain tasks.

While recent advances in LLM-based cybersecurity research have demonstrated notable improvements, most existing approaches remain limited to specific tasks or require substantial expert involvement. Different from these works, LLM-PD integrates LLMs into a comprehensive proactive defense pipeline. By leveraging prompt-driven reasoning, modular agent collaboration, strict action validation, and memory-based feedback, we enable adaptive defense decisions and rapid deployment of mitigation mechanisms in the dynamic cloud environment.

\section{Threat Model}
We consider an adversary capable of launching a variety of DoS attacks, including SYN flooding, SlowHTTP, and Memory DoS, against cloud services. The attacker is assumed to have knowledge of standard network protocols and can generate malicious traffic or resource contention to disrupt service availability. However, the attacker does not have privileged access to the cloud infrastructure or the defense system itself. The defender operates under the assumption that attack patterns may be adaptive and evolving, but the underlying cloud management and monitoring infrastructure remains trustworthy.

\begin{figure*}[t]
  \centering
  \includegraphics[width=\textwidth]{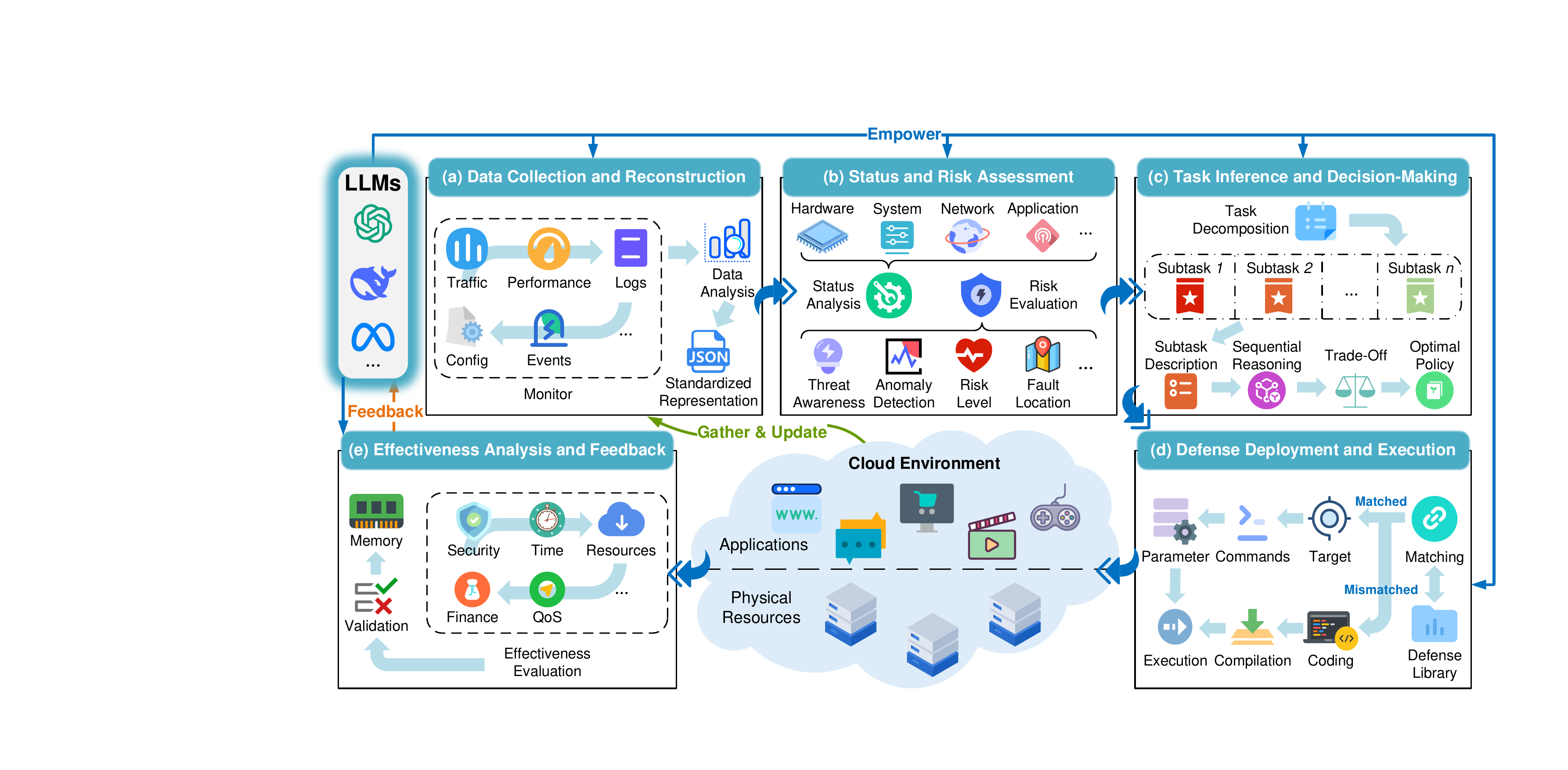}
  \caption{
    An overview of the proposed LLM empowered proactive defense architecture in cloud networks. The architecture consists of five main LLM-driven agents, each responsible for a key stage in the proactive defense lifecycle: (a) \textbf{Data Collector} gathers heterogeneous security data from the cloud environment; (b) \textbf{Analyzer} assesses system status and quantifies risks; (c) \textbf{Decision-Maker} decomposes defense tasks and generates optimal strategies; (d) \textbf{Deployer} executes or generates defense actions and scripts; (e) \textbf{Feedback Giver} evaluates defense effectiveness and updates memory for continuous improvement. The agents interact in a closed feedback loop, enabling autonomous, adaptive, and evolving cloud security defense.
  }
  \label{architecture}
\end{figure*}

To achieve autonomous, AI-driven, and secure-by-design defense, LLM-PD is architected to ensure robustness through continuous system monitoring, dynamic adaptation of defense strategies, and memory-based feedback that prevents the repetition of ineffective actions, as illustrated in Fig.~\ref{architecture}. Even in adversarial environments characterized by unknown or adaptive attack strategies, the architecture rigorously validates all defense actions, systematically filters out ineffective or unsafe responses, and leverages experiential learning from previous episodes to enhance resilience over time.

\section{System Architecture}

\subsection{Data Collection and Reconstruction}
The data collector agent is responsible for gathering, integrating, and preprocessing heterogeneous security data from various sources in the cloud environment, providing a unified and structured input for subsequent analysis.

There exist a substantial volume of data in cloud networks, including network traffic, performance metrics, logs, events, and service configurations, etc. Essentially, each type of data is considered independent and is formatted differently. The ability to proficiently use multiple security tools and analyze massive heterogeneous data poses a significant challenge, even for cybersecurity professionals.

Different from typical approaches that rely on fixed schemas or manual parsing, our collector implements semantically-driven heterogeneous data integration that can flexibly interpret and normalize diverse, unstructured data formats, enabling rapid adaptation to new sources. Nevertheless, given the large number of data sources, redundant and irrelevant information is inevitable. To enable prompt-based LLM analysis, the collector applies log parsing, normalization, deduplication, and semantic aggregation. Raw data from multiple sources are parsed and mapped to extract relevant fields. For example, multiple logs or alerts may show as: \texttt{\footnotesize\{"timestamp": "2025-06-01T10:00:00Z", "event": "failed\_login", "source\_ip": "192.168.1.10", "username": "admin", "status": "failure"\}, \{"timestamp": "2025-06-01T10:00:01Z", "event": "failed\_login", "source\_ip": "192.168.1.10", "username": "admin", "status": "failure"\}, \{"timestamp": "2025-06-01T10:00:02Z", "event": "heartbeat", "status": "ok"\}}. The collector can aggregate the repeated failed login attempts into a summarized alert and filter out the routine heartbeat message as irrelevant to the current security analysis.

Finally, this module reconstructs the refined results into a standardized representation (e.g., JSON file) and transmits it to the next stage for further processing.

\subsection{Status and Risk Assessment}
The analyzer agent evaluates the collected data to assess the current system status, identify potential security risks, enabling informed and prioritized defense planning.

\subsubsection{Status Analysis}
Within the status analysis function, information regarding hardware (e.g., power consumption), system (e.g., system load), network (e.g., traffic volumes), and applications (e.g., number of connections) can be extracted from standardized data fields. This status information not only reflects the operational state but also allows for the derivation of constraints for subsequent defense tasks. For example, when the system utilization is high, the analyzer recognizes resource constraints and dynamically prioritizes and allocates available resources to ensure that critical defense objectives are met. In the event of critical threats, the system is designed to guarantee that essential defensive actions are executed with the highest priority, even under high load, by preempting non-critical tasks and reallocating resources as needed.

\subsubsection{Risk Evaluation}
This function analyzes security indicators from monitoring tools and historical data to identify potential threats. Concurrently, the risk level is quantified based on three main factors: (i) Scope represents the extent of affected resources or services (e.g., number of impacted hosts, services, or users). (ii) Impact reflects the severity of the potential or observed consequences (e.g., service interruption, data loss, or performance degradation). (iii) Duration indicates the length of time the threat has persisted or is likely to persist. Each factor is mapped to a sub-score (e.g., $0–3$ for scope, $0–4$ for impact, $0–3$ for duration), and the total risk score is the sum of these sub-scores, normalized to a $0–10$ scale. This quantitative risk score is then used to prioritize defense actions and allocate resources efficiently.

Unlike rule-based analyzers that rely on static rules or require extensive retraining for new scenarios, we leverage the reasoning and generalization capabilities of LLMs to establish cause-effect relationships between attack indicators and system anomalies. The LLM agent is prompted with structured historical and real-time data, and required to output both inferred causal links and a brief explanation of its reasoning process. For unknown attack indicators and anomalies, the analyzer utilizes the LLM's ability to recognize abnormal patterns, such as sudden deviations from historical baselines, allowing adaptive detection of novel threats beyond predefined rules.

\subsection{Task Inference and Decision-Making}
In complex cloud environments, however, multiple threats or failure points may arise simultaneously, requiring the simultaneous achievement of multiple defense objectives within a single overarching task. Nevertheless, the complexity of these tasks can lead to conflicts, when conventional approaches that rely on predefined logic are employed. To prevent task conflicts and facilitate efficient decision-making, we have developed a hierarchical decision-maker agent that infers which defense tasks should be performed, decomposes complex tasks into subtasks, and generates optimal defense strategies accompanied by explicit rationales.

\subsubsection{Task Decomposition}
In a high-level threat scenario involving multiple defense objectives, it is crucial to clearly plan and generate detailed tasks. To this end, we have designed a task decomposition function that breaks down complex tasks into independent subtasks. We require the LLM to delineate the implementation constraints of each subtask, assign priorities based on the previously mentioned risk levels, and analyze the dependencies among tasks for sequential arrangement.

\subsubsection{Inference and Decision-Making}
In this process, tasks are executed sequentially according to their assigned order. For each subtask, the function systematically extracts task details and, considering current requirements and constraints, performs human-like sequential reasoning to determine the necessary defensive actions. To mitigate the risk of LLM hallucinations (i.e., generating invalid or unreasonable defense actions), the decision-maker incorporates strict parameter validation and environmental constraint checks before generating any policy. Actions with parameters outside valid ranges or violating resource constraints are automatically rejected. In addition, it employs an exploration rate mechanism to balance the exploitation of historically successful strategies and the exploration of new action sequences, further enhancing robustness and reducing the risk of repeated hallucinated decisions. Finally, by balancing multiple competing objectives, it imports solvers to optimize the policy among the available strategies.

\subsection{Defense Deployment and Execution}
Unlike conventional approaches that rely on manually crafted scripts or static automation, the LLM agent can autonomously generate, validate, and deploy customized defense code in response to threats and dynamic requirements, greatly expanding the adaptability and coverage of the system without extensive human intervention.
Upon receiving a defense strategy, the deployer queries the matching function to check for corresponding items in the defense database. If a match is found, it extracts deployment targets, commands, and parameters, then invokes the appropriate defense actions to align with strategic objectives.

When a required mechanism is absent from the defense library, it leverages LLMs to generate scripts via prompt-based code generation, constructing the necessary function on-the-fly. The process is as follows: (i) The LLM receives a natural language prompt describing the defense objective, target environment, and constraints, and then, it generates the corresponding code or script. (ii) The generated code is automatically validated through syntax checking, static analysis, and sandboxed test execution, ensuring that hallucinated or unsafe outputs are filtered out before affecting the system. (iii) Upon successful verification, the script is deployed to the target system for execution. (iv) The new script, along with its metadata (e.g., purpose, environment, effectiveness), is archived in the defense library for future reuse.

For example, given the input prompt:
\begin{lstlisting}[basicstyle=\ttfamily\footnotesize,breaklines=true,breakindent=0pt]
Generate a Bash script to block all incoming traffic from IP 192.168.1.10, and log this action to /var/log/defense.log.
\end{lstlisting}
		
The LLM may output:
\begin{lstlisting}[basicstyle=\ttfamily\footnotesize,breaklines=true,breakindent=0pt]
#!/bin/bash
IP="192.168.1.10"
iptables -A INPUT -s $IP -j DROP
echo "$(date): Blocked all incoming traffic from $IP" >> /var/log/defense.log
\end{lstlisting}

\ The significance of this multi-stage validation process is that it ensures that all autonomously generated solutions are reliable and traceable, expanding the adaptability and coverage of the system, even in the absence of human oversight.

\subsection{Effectiveness Analysis and Feedback}
Existing approaches in cloud security are largely static, lacking the ability to adjust and evolve in response to new threats or changing environments. In contrast, a key innovation of LLM-PD lies in its memory feedback mechanism, which enables the system to learn and adapt over time. This agent evaluates the effectiveness of deployed defense actions using multi-dimensional metrics, verifies if they meet expectations, records the annotated defense sequences, and updates the memory to enable continuous learning. By referencing this historical experience, LLM-PD can avoid previously ineffective actions, reinforce successful strategies, and dynamically refine its decision-making process without retraining.

\subsubsection{Multi-dimensional Evaluation}
LLM-PD evaluates defense effectiveness using a multi-dimensional approach that integrates security effectiveness, recovery time, resource consumption, financial cost, and Quality of Service (QoS), enabling a comprehensive assessment beyond a single security metric. Each dimension is assigned an importance weight based on the specific defense context and organizational priorities, enabling dynamic trade-offs that balance robust protection, minimize operational overhead, and optimize user experience throughout the defense process.

\subsubsection{Endpoint Validation}
This function determines whether the current defense process has concluded. After each defensive action, it evaluates the system state against successful mitigation and failure conditions. Once the round ends, it tags the sequence of actions with outcome flags (success or failure) and relevant contextual metadata, then transmits this annotated sequence to the memory module for future reference.

\subsubsection{Memory-Based Optimization}
The memory module implements a dual-layer memory architecture that consists of intra-episode pool and inter-episode pool. The intra-episode memory pool records stepwise defense actions and outcomes within a single episode, while the inter-episode memory pool employs a sliding window mechanism to retain only the most recent batches of complete defense action sequences and their validation results. This closed-loop feedback mechanism not only addresses token limit constraints but also reinforces successful strategies and avoids previously ineffective action sequences. By maintaining a memory of successful and failed decisions, validation results, and decision rationales, the system proactively prevents the repetition of hallucinated or invalid actions in future episodes, thereby facilitating continuous improvement across multiple defense episodes.

\section{Case Study}
\subsection{Experimental Setups}
It's important to note that our architecture is model-agnostic and can be implemented with various LLMs. For our current implementation, we simulate a prototype of LLM-PD with GPT-4o mini, DeepSeek-R1-Distill-Qwen-32B, and Qwen3-32B, respectively. We configure GPT-4o mini with temperature 1.0 and Top-$p$ 1.0, DeepSeek-R1-Distill-Qwen-32B with temperature 0.6 and Top-$p$ 0.95, and Qwen3-32B with temperature 0.7 and Top-$p$ 0.8 to optimize their respective performance. Table~\ref{prompts} presents the example prompts for each component.

\begin{table}[t]
  \caption{Prompts for each component in the proposed architecture.}
  \centering
  \label{prompts}
  \renewcommand{\arraystretch}{1.25}
  \begin{tabular}{@{}cm{2.59in}@{}}
  \toprule
  \textbf{Components} & \multicolumn{1}{c}{\textbf{Examples of Prompts}} \\ \arrayrulecolor{black} \midrule
  Basic Profile & You are a security robot capable of continuously improving defense strategies across multiple [\textit{Episodes}] of DoS attacks. Each episode consists of multiple [\textit{Steps}] in the attack and defense processes. You must constantly monitor the cloud networks to ensure system security and service availability. \\ \arrayrulecolor{lightgray} \hline
  Collector & You are responsible for collecting heterogeneous security data from the cloud environment. At each step, you may call optional tools [\textit{Wireshark, Tcpdump, ...}] to capture network traffic, or parse system logs and configuration files. If you detect repeated or irrelevant events in the logs, aggregate them and output the cleaned, structured data in [\textit{JSON, XML, ...}] format. \\\arrayrulecolor{lightgray} \hline
  Analyzer & Given the current system status, including [\textit{CPU Usage, Memory Usage, Number of Connections, and Detected Anomalies}], evaluate the risk level by considering the [\textit{Scope, Impact, and Duration}] and output a normalized risk score (0–10) with a brief explanation. \\\arrayrulecolor{lightgray} \hline
  Decision-Maker & Given the [\textit{Priorities}] and current [\textit{Risk Score}], decompose the overall defense task into subtasks. For each subtask, select the most suitable defense action from the available options [\textit{MTD mechanisms, cyber deception methods, ...}]. By considering [\textit{Constraints}] and [\textit{Preferences}], output the recommended action sequence and rationale. \\\arrayrulecolor{lightgray} \hline
  Deployer & For each recommended [\textit{Action}] to be taken in this round, if an existing script or mechanism is available, retrieve and execute it with the specified [\textit{Configurations}]. If not, generate a new script to implement the action, validate its correctness, and deploy it. \\\arrayrulecolor{lightgray} \hline
  Feedback Giver & Please evaluate the defense effectiveness based on [\textit{Security, Time, ...}] and the [\textit{Validity}] of strategy with its implementation. Record whether the defense was successful, the reasons for any failures, and lessons learned. Store this feedback in [\textit{Memory}] for future decision-making and strategy refinement. \\ \arrayrulecolor{black}
  \bottomrule
  \end{tabular}
\end{table}

To evaluate our proposed architecture in defeating threats, we implement a DoS attack scenario. Specifically, for SYN flooding and SlowHTTP attacks, we deploy two attacker instances, each assigned a distinct IP address to simulate independent external attack sources. We simulate an elastic service that can create up to 10 replicas, using a total pool of 100 pods. Each pod supports 256 connections and a maximum memory utilization of 100. The initial setup activates 5 replicas, each comprising 10 pods. Additionally, Memory DoS is modeled as a co-resident attack, where attacker processes are launched within the same physical infrastructure to induce resource contention among Virtual Machines (VMs). We model a cluster with 5 racks, each containing 10 physical machines, and each machine capable of hosting up to 10 VMs. The maximum number of memory contention per VM is limited to 100 with runtime capped at 100 time steps.

In our experimental setup, each defense module operates in discrete time steps, with each step corresponding to 30 seconds of simulated time. A complete cycle through all five modules constitutes a single round. An episode is defined as the duration of a single attack scenario, typically consisting of multiple rounds. The episode continues until the system reaches a stable state, either secure or compromised for 5 consecutive rounds. This setup allows us to observe the iterative and adaptive nature of LLM-PD, as the system continuously refines its defense strategies within each episode based on real-time feedback and evolving attack conditions.

We compare our method with DQN~\cite{zhou2025resource}, Actor-Critic (AC)~\cite{zhang2023disturb}, and Proximal Policy Optimization (PPO)~\cite{zhang2023towards}. The discount factor is set to 0.98, with a learning rate of 0.001 for the Q-network in DQN and policy networks in AC and PPO, and 0.01 for value networks in AC and PPO. All networks have a single hidden layer of 256 neurons. For statistical reliability, we conduct 200 independent tests, each with 10 episodes.

\subsection{Results}
\begin{figure}[t]
  \centering
  \includegraphics[width=\columnwidth]{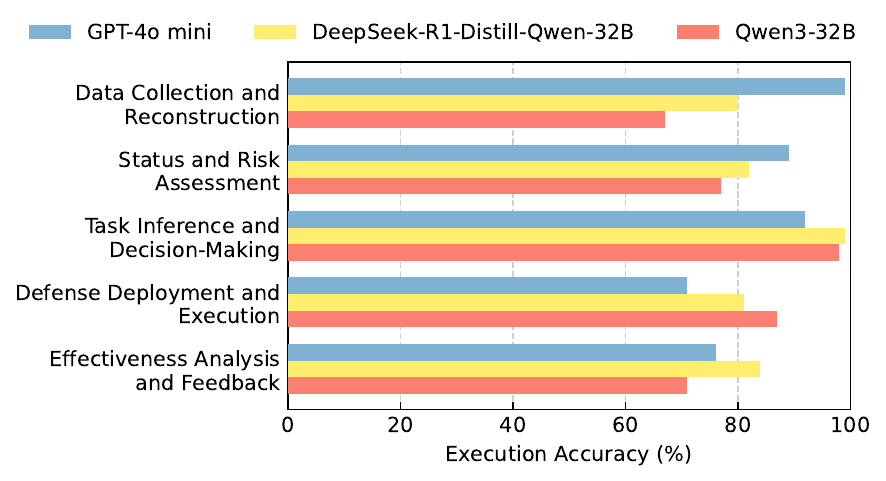}
  \caption{Execution accuracy of LLM-PD across five defense pipeline stages using different LLMs. All models maintain high accuracy, with each exhibiting unique strengths across different defense stages.}
  \label{execution_accuracy}
\end{figure}

\begin{table}[t]
  \centering
  \caption{LLM-PD achieves consistently high surviving rates across all DoS attack scenarios, with DeepSeek-R1-Distill-Qwen-32B and Qwen3-32B showing the best performance and stability.}
  \label{surviving_rates}
  \begin{tabular}{@{}lccc@{}}
    \toprule
    \textbf{DoS Attack} & \textbf{GPT-4o mini} & \textbf{\begin{tabular}[c]{@{}c@{}}DeepSeek-R1-Distill-\\ Qwen-32B\end{tabular}} & \textbf{Qwen3-32B} \\ \midrule
    SYN Flooding & 0.888 $\pm$ 0.092 & 0.961 $\pm$ 0.057 & 0.977 $\pm$ 0.026 \\
    SlowHTTP     & 0.918 $\pm$ 0.080 & 0.985 $\pm$ 0.035 & 0.982 $\pm$ 0.020 \\
    Memory DoS   & 0.936 $\pm$ 0.061 & 0.931 $\pm$ 0.104 & 0.936 $\pm$ 0.086 \\ \bottomrule
    \end{tabular}
  \end{table}

To evaluate our architecture, we measure the execution accuracy for each stage in the defense pipeline in Fig.~\ref{execution_accuracy}, calculated as the percentage of successful outcomes per trial. The results reveal two key insights. First, the architecture is fundamentally robust, as all models achieve high overall accuracy. Second, and more importantly, the performance nuances highlight that different LLMs exhibit distinct aptitudes for specific tasks. For instance, GPT-4o mini excels in the initial stages, achieving 99\% in data collection and 89\% in risk assessment. In contrast, DeepSeek-R1-Distill-Qwen-32B and Qwen3-32B demonstrate superior capabilities in the decision-making stage, with accuracies of 99\% and 98\% respectively. Furthermore, Qwen3-32B shows the highest proficiency of 87\% in defense deployment, while DeepSeek-R1-Distill-Qwen-32B is the most effective in the feedback stage. This validates that while our architecture is effective by design, performance can be further optimized by selecting the most suitable LLM for the most critical phases of the defense lifecycle.

We present a comparative analysis of LLM-PD when implemented with different LLMs in Table~\ref{surviving_rates}. In this context, the surviving rate refers to the proportion of time steps during which the system successfully withstands ongoing attacks without service disruption, after a defensive action is taken. To ensure statistical reliability, we report the mean surviving rate and its 95\% confidence interval. Specifically, while all models perform effectively, we observe performance nuances among them. DeepSeek-R1-Distill-Qwen-32B and Qwen3-32B generally exhibit higher surviving rates with tighter confidence intervals compared to GPT-4o mini in defeating SYN Flooding and SlowHTTP attacks. Nevertheless, GPT-4o mini maintains a commendable performance, particularly in the Memory DoS scenario, where it matches the surviving rate of Qwen3-32B. This suggests that the reasoning and inference capabilities of the underlying model can further enhance defense precision and consistency, with different models demonstrating varying suitability for different attack scenarios.

\begin{figure}[t]
  \centering
  \includegraphics[width=0.85\columnwidth]{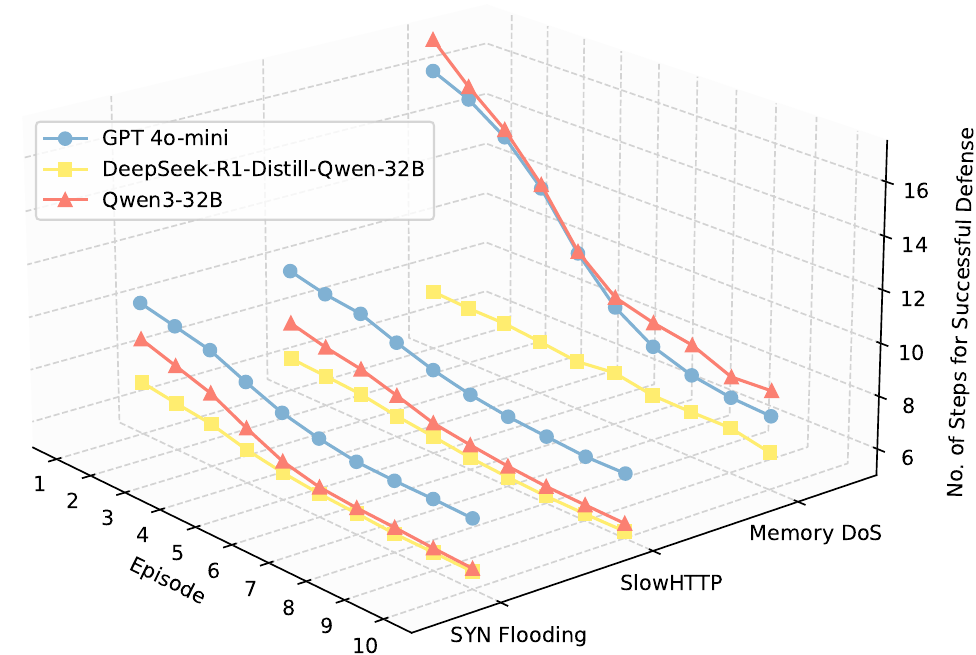}
  \caption{LLM-PD rapidly reduces the number of steps needed for successful defense as episodes progress, demonstrating effective self-evolution and learning across all attack types.}
  \label{steps}
\end{figure}

Fig.~\ref{steps} illustrates the learning efficiency of our architecture that compares three LLM implementations. This figure maps learning episodes on the $x$-axis, attack scenarios on the $y$-axis, and the average defense steps required for successful defense on the $z$-axis. A clear observation is that more steps are required in initial episodes, reflecting the varying difficulty of each attack when the system has limited experience. For instance, using GPT-4o mini in the first episode requires 10.05, 9.69, and 16.02 steps for SYN Flooding, SlowHTTP, and Memory DoS, respectively. As episodes progress, a consistent downward trend demonstrates the effectiveness of the self-evolution mechanism. By the 10$th$ episode, for example, the steps for the DeepSeek-R1-Distill-Qwen-32B model converge to 5.64, 5.26, and 6.40 for the respective attacks. This improvement is attributed to LLM-PD's memory feedback loop, which facilitates learning from past interactions to avoid ineffective strategies and promote efficient threat mitigation.

In addition, we compare LLM-PD against three baseline solutions in Table~\ref{performance_comparison}, evaluating three key metrics, including defense efficacy (the success rate of mitigating attacks), latency (the time taken to execute a defense action), and cost (the financial expense incurred per defense episode). The results show that our LLM-PD architecture consistently achieves the highest defense efficacy across all attack types. For instance, in the SYN Flooding scenario, all LLM-PD variants surpass 97\% efficacy, decisively outperforming the best baseline (DQN at 82.68\%) and demonstrating superior adaptability. However, this enhanced efficacy comes at the cost of higher latency and resource consumption, an expected trade-off for leveraging LLMs. Nevertheless, the ability of LLM-PD to generalize across different attack vectors without specific retraining highlights a crucial advantage over traditional methods. This implies that for critical systems where security is paramount, the investment in higher latency and cost yields a more robust and reliable defense posture.

\begin{table*}[t]
  \centering
  \begin{threeparttable}
    \caption{LLM-PD outperforms baseline methods in defense efficacy across all attack scenarios, at the cost of higher latency and resource consumption, highlighting a trade-off between robustness and efficiency.}
    \label{performance_comparison}
    \begin{tabular}{@{}l *{3}{ccc}@{}}
      \toprule
      \multirow{2}{*}{\textbf{Method}} & \multicolumn{3}{c}{\textbf{SYN Flooding}} & \multicolumn{3}{c}{\textbf{SlowHTTP}} & \multicolumn{3}{c}{\textbf{Memory DoS}} \\ \cmidrule(l){2-4} \cmidrule(l){5-7} \cmidrule(l){8-10}
      & \textbf{Efficacy (\%)} & \textbf{Latency (s)} & \textbf{Cost (\$)} & \textbf{Efficacy (\%)} & \textbf{Latency (s)} & \textbf{Cost (\$)} & \textbf{Efficacy (\%)} & \textbf{Latency (s)} & \textbf{Cost (\$)} \\ \midrule
      DQN~\cite{zhou2025resource} & 82.68 & 0.0595 & 0.0135 & 84.94 & 0.0601 & 0.0136 & 56.07 & 0.2117 & 0.0479 \\
      AC~\cite{zhang2023disturb} & 73.65 & 0.1189 & 0.0269 & 77.35 & 0.1134 & 0.0256 & 54.90 & 0.6142 & 0.1389 \\
      PPO~\cite{zhang2023towards} & 73.00 & 0.5429 & 0.1228 & 78.79 & 0.4740 & 0.1072 & 55.56 & 1.1341 & 0.2565 \\
      LLM-PD\tnote{*} & 97.08 & 2.7255 & 3.2045 & 98.47 & 2.6646 & 3.3227 & 98.78 & 2.5523 & 2.3432 \\
      LLM-PD\tnote{\dag} & 98.15 & 7.1682 & 0.0418 & 98.02 & 7.3874 & 0.0322 & 98.88 & 7.3286 & 0.0301 \\
      LLM-PD\tnote{\ddag} & 98.66 & 12.411 & 0.0492 & 97.69 & 13.136 & 0.0258 & 93.87 & 16.394 & 0.0809 \\ \bottomrule
    \end{tabular}
    \begin{tablenotes}[para,flushleft]
      \footnotesize
      \item *: \textit{w/} GPT-4o mini;
      \item \dag: \textit{w/} DeepSeek-R1-Distill-Qwen-32B;
      \item \ddag: \textit{w/} Qwen3-32B.
    \end{tablenotes}
  \end{threeparttable}
\end{table*}

\section{Conclusion and Future Work}
The development of LLMs is promising for tackling challenges associated with mitigating cyberattacks. In this paper, we introduce LLM-PD, a novel multi-agent architecture for proactive DoS defense in cloud environments. By leveraging the advanced capabilities of LLM in data collection, security analysis, task inference, defense deployment, and effectiveness evaluation, our method is capable of thoroughly analyzing the security situation, efficiently executing suitable actions, and continuously evolving itself to adapt to varying DoS attack scenarios. We then present a detailed case study on three types of DoS attacks. Experimental results demonstrate that the proposed architecture improves the effectiveness and efficiency of defense compared to state-of-the-art methods.

While our current evaluation focuses on synthetic DoS scenarios, these experiments only reflect a subset of the potential application domains. Importantly, the architecture of LLM-PD is inherently designed with strong extensibility and adaptability in mind. By leveraging prompt-driven task inference, modular agent design, and memory-based self-evolution, LLM-PD can be readily tailored to address a wide variety of threat types and operational contexts. In future work, we plan to extend our experiments to more complex and realistic cloud environments, including APTs, supply chain attacks, and multi-stage attack campaigns. With appropriate prompt engineering and agent optimization, LLM-PD can adapt its defense strategies and workflows to these sophisticated scenarios.

\section*{Acknowledgments}
This work was supported in part by the National Natural Science Foundation of China under Grant No. 62202097 and Grant No. 62072100, in part by the Frontier Technologies R \& D Program of Jiangsu under Grant No. BF2025026, in part by China Postdoctoral Science Foundation under Grant No. 2024T170143 and No. 2022M710677, and in part by Jiangsu Funding Program for Excellent Postdoctoral Talent under Grant No. 2022ZB137.

%
\bibliographystyle{IEEEtran}
\bibliography{IEEEabrv,Reference}


\section*{Biographies}
\vspace{-28pt}
\begin{IEEEbiographynophoto}{Yuyang Zhou} [M'22] (yyzhou@seu.edu.cn) is currently working as a postdoc with the School of Cyber Science and Engineering, Southeast University.
\end{IEEEbiographynophoto}
\vspace{-28pt}
\begin{IEEEbiographynophoto}{Guang Cheng} [M'04] (chengguang@seu.edu.cn) is a Full Professor in the School of Cyber Science and Engineering, Southeast University.
\end{IEEEbiographynophoto}
\vspace{-28pt}
\begin{IEEEbiographynophoto}{Kang Du} (dukang@seu.edu.cn) is currently pursuing his master degree with the School of Cyber Science and Engineering at Southeast University.
\end{IEEEbiographynophoto}
\vspace{-28pt}
\begin{IEEEbiographynophoto}{Zihan Chen} [M'23] (zhchen@njnet.edu.cn) is a postdoc researcher with the School of Cyber Science and Engineering at Southeast University.
\end{IEEEbiographynophoto}
\vspace{-28pt}
\begin{IEEEbiographynophoto}{Yuyu Zhao} [M'23] (yyzhao@seu.edu.cn) is a lecturer with the School of Cyber Science and Engineering at Southeast University.
\end{IEEEbiographynophoto}
\vfill
\end{document}